\def\half{\frac{1}{2}}
\def\mb#1{\mbox{\boldmath{$#1$}}} 

\documentclass{iopart}
\usepackage{cite}
\begin{document}
\hspace*{4.5 in}CUQM-137 \\

\vskip 0.4 in


\title{Generalized quantum isotonic nonlinear oscillator in $d$ dimensions}


\author{Richard L. Hall$^1$, Nasser Saad$^2$, and \"Ozlem Ye\c{s}ilta\c{s}$^3$}

\address{$^1$Department of Mathematics and Statistics, Concordia
University, 1455 de Maisonneuve Boulevard West, Montr\'eal,
Qu\'ebec, Canada H3G 1M8. }

\address{$^2$ Department of Mathematics and Statistics,
University of Prince Edward Island, 550 University Avenue,
Charlottetown, PEI, Canada C1A 4P3.}

\address{$^3$Department of Physics, Faculty of Arts and Sciences,
Gazi University, 06500 Ankara, Turkey.}

\ead{$^1$rhall@mathstat.concordia.ca, $^2$nsaad@upei.ca, $^3$yesiltas@gazi.edu.tr}

\begin{abstract}
We present a supersymmetric analysis for the $d$-dimensional Schr\"{o}dinger equation with the generalized isotonic nonlinear-oscillator potential $V(r)={B^2}/{r^{2}}+\omega^{2} r^{2}+2g{(r^{2}-a^{2})}/{(r^{2}+a^{2})^{2}}$, $B\geq 0$. We show that the eigenequation for this potential is exactly
 solvable provided $g=2$ and $(\omega a^2)^2 = B^2 +(\ell +(d-2)/2)^2.$ Under these conditions, we obtain explicit formulae for all the  energies and normalized bound-state wavefunctions.
\end{abstract}
\maketitle
\section{Introduction}
Recently, Cari\~nena \emph{et al.} \cite{perelomov} studied Schr\"odinger's equation with a quantum nonlinear oscillator potential of the form
\begin{equation}\label{eq1}
\left[-{d^2\over dx^2}+x^2+8{2x^2-1\over (2x^2+1)^2}\right]\psi_n(x)=E_n\psi_n(x).
\end{equation}
Their interest in this study came from the fact that Eq.(\ref{eq1}), is exactly solvable, in a sense that the eigenenergies and eigenfunctions can be obtained explicitly; they were able to show that \cite{perelomov}
\begin{equation}\label{eq2}
\left\{ \begin{array}{l}
 \psi_n(x)={P_n(x)\over (2x^2+1)}e^{-x^2/2},\\ \\
  E_n=-3+2n,\quad n=0,3,4,5,\dots,
       \end{array} \right.
\end{equation}
where the polynomial factors $P_n(x)$ are related to the Hermite polynomials by
\begin{equation}\label{eq3}
P_n(x)=\left\{ \begin{array}{ll}
 1 &\mbox{ if $n=0$} \\ \\
  H_n(x)+4nH_{n-2}(x)+4n(n-3)H_{n-4}(x) &\mbox{ if $n=3,4,5,\dots$}
       \end{array} \right.
\end{equation}
Soon afterwards,  Fellows and Smith \cite{fellows} considered another interesting case and, in particular, they showed that for certain values of the parameters $w$ and $g$, the potential in the Schr\"odinger equation
\begin{equation}\label{eq4}
\left[-{d^2\over dx^2}+w^2x^2+2g{x^2-a^2\over (x^2+a^2)^2}\right]\psi_n(x)=2E_n\psi_n(x)
\end{equation}
is indeed a supersymmetric partner of the harmonic oscillator potential. By means of the supersymmetric approach, Fellows and Smith \cite{fellows} were able to construct an infinite set of exactly soluble \emph{partner} potentials, along with their eigenfunctions and eigenvalues. Very recently, Sesma \cite{sesma}, using a M\"obius transformation, was able to transform Eq.(\ref{eq4}) into a confluent Heun equation and thereby obtain a simple and efficient algorithm to solve the Schr\"odinger equation (\ref{eq4}) numerically, no matter what values are chosen for the parameters $w$ and $g$. Furthermore, using suitable mass distributions, the position-dependent effective mass Schr\"{o}dinger equation has been solved for a new nonlinear oscillator \cite{kra}. We  note that the term $2(x^2-a^2)/(x^2+a^2)^2$ in Eq.(\ref{eq4}) can be written as the sum of two centripetal barriers in the complex plane
$${2(x^2-a^2)\over (x^2+a^2)^2}={1\over (x+\imath a)^2}+{1\over (x-\imath a)^2}, \quad \imath=\sqrt{-1},$$
that is to say, a rational potential with two imaginary poles symmetrically placed with respect to the origin \cite{perelomov}.

\vskip0.1true in
\noindent The purpose of the present work is to use the supersymmetric approach to analyze the exact \emph{analytic} solutions for the more general potential class (\ref{eq4}).  We consider the $d-$dimensional Schr\"odinger eigenvalue problem
\begin{equation}\label{eq5}
H\Psi(\mb{r}) = E\Psi(\mb{r}),\quad H = -\Delta +V,
\end{equation}
where $\Delta$ is the $d-$dimensional Laplacian operator, $d\ge 2$, and $V(r)$ is the central potential  given by
\begin{equation}\label{eq6}
V(r) = \frac{B^2}{r^2} + \omega^2 r^2 + 2g\frac{(r^2-a^2)}{(r^2+a^2)^2},\quad B^2\geq 0.
\end{equation}
We show, through the factorization method, that the eigenequation for this potential, Eq.(\ref{eq5}), is indeed exactly
solvable provided $g=2$ and $(\omega a^2)^2 = B^2 +(\ell +(d-2)/2)^2.$ The article is organized as follows. In section 2, we consider the $d$-dimensional Schr\"odinger equation with the nonlinear oscillator potential (\ref{eq6}). In section 3, we briefly review the factorization method and the basic fomulae from supersymmetric theory that we need for our investigation \cite{cooper,hall2}.  In section~4, we show that the isotonic nonlinear-oscillator potential \cite{fellows, perelomov} and the Gol'dman and Krivchenkov potential are isospectral supersymmetric partner potentials. In section~6, the explicit construction of all the exact solutions of Schr\"odinger's equation (\ref{eq5}) with $V(r)$ given by (\ref{eq6}) is presented.
\section{Schr\"{o}dinger's equation  in $d$ dimensions}
By considering the action of the Laplacian operator on a wavefunction of the form $\Psi(\mb{r})=u(r)Y_l(\theta_1,\theta_2,\dots,\theta_{d-1})$ with a spherically symmetric factor $u(r)$ and a generalized spherical harmonic factor $Y_l$, we  obtain the radial Schr\"{o}dinger equation in $d$ dimensions for a spherically symmetric potential $V(r)$ as \cite{sommerfeld, hall1}:
\begin{equation}\label{eq7}
    -\frac{d^{2}u}{dr^{2}}-\frac{d-1}{r}\frac{du}{dr}+\frac{\ell(\ell+d-2)}{r^{2}}u+V(r)u=Eu,
\end{equation}
where, if $\mb{r}\in\Re^d$, then $r = \|\mb{r}\|,$ and
\[
u(r)\in L^{2}([0,\infty),r^{d-1}dr).
\]
The first-order derivative term can be removed  by using the new radial function $\psi(r)$ given by
\[
\psi(r)=r^{\frac{d-1}{2}} u(r),\quad d\geq 2, \quad \psi(0)=0,\quad \psi(r)\in L^{2}([0,\infty),dr).
\]
Thus, we find
\begin{equation}\label{eq8}
   -\frac{d^{2}\psi}{dr^{2}}  + U(r) \psi=E \psi
\end{equation}
where
\begin{equation}\label{eq9}
    U(r)=V(r)+\frac{(2\ell+d-1)(2\ell+d-3)}{4r^{2}}.
\end{equation}
We  now recall the definition of $V(r)$ in (\ref{eq6}) and combine the terms in $1/r^2$ to obtain,
\begin{equation}\label{potU}
U(r) = \frac{k(k+1)}{r^2} + \omega^2 r^2 + 2g\frac{(r^2-a^2)}{(r^2+a^2)^2},
\end{equation}
where $k$ is defined so that
\[
B^2 + \frac{1}{4}(2\ell+d-1)(2\ell+d-3) = k(k+1),
\]
that is to say,
\begin{equation}\label{defL}
k = \left[B^2 +\left(\ell + \frac{d-2}{2}\right)^2\right]^{\half} - \half.
\end{equation}
We shall find it convenient to label energy eigenstates $\psi_{nk} \equiv \psi_{n\gamma_d}$, where $n = 0,1,2,\dots$ is the number of radial nodes, and $\gamma_d = k+\half.$
\section{The factorization method}
In this section we give a brief review of some concepts of supersymmetric quantum mechanics (SUSY QM) that we shall need  in the following sections, namely the factorization method, and supersymmetric partner potentials. We start with Schr\"odinger's time-independent equation for a one-dimensional radial problem (in the units $\hbar=2m=1$)
\begin{equation}\label{eq12}
    H\psi(r)=\left[-\frac{d^{2}}{dr^{2}}+U(r)\right]\psi(r)=E\psi(r)
\end{equation}
where the potential $U(r)$ is real and possibly singular at $r=0.$ The wave function $\psi(r)$ must be square-integrable and normalized in a sense that $\int |\psi(r)|^2dr=1$. The main idea of the factorization method, as introduced in this context by Schr\"odinger and Dirac,
is to write the second order differential operator $H$ in (\ref{eq12}) as the product of two first
order differential operators $\mathcal{A}^{\dag}$ and $\mathcal{A}$, such that $H=\mathcal{A}^{\dag}\mathcal{A}$ and
\begin{equation}\label{eq13}
   \mathcal{A}^{\dag}=-\frac{d}{dr}+W(r),\quad  \mathcal{A}=\frac{d}{dr}+W(r)
\end{equation}
where the function $W(r)$ is a real function of $r$ and is known as the superpotential of the problem. The operators $\mathcal{A}$ and $\mathcal{A}^{\dag}$ are Hermitian conjugates of each other: $(\mathcal{A}^{\dag})^{\dag}=\mathcal{A}$.
Let us define
\begin{equation}\label{eq14}
H_1=\mathcal{A}^{\dag} \mathcal{A}=-\frac{d^{2}}{dr^{2}}+U_{1}(r), ~~~~H_2=\mathcal{A}\mathcal{A}^{\dag} =-\frac{d^{2}}{dr^{2}}+U_{2}(r).
\end{equation}
In general $H_1\equiv H$ and $H_2$ are two different Hamiltonians. They are known as partner Hamiltonians in SUSY QM and are given explicitly by:
\begin{eqnarray}\label{eq15}
H_1&=&\mathcal{A}^{\dag} \mathcal{A}=\left(-\frac{d}{dr}+W(r)\right)\left(\frac{d}{dr}+W(r)\right)=-\frac{d^2}{dr^2}+W^2(r)-{dW(r)\over dr}=-\frac{d^2}{dr^2}+U_1(r)
\end{eqnarray}
and
\begin{eqnarray}\label{eq16}
H_2&=&\mathcal{A}\mathcal{A}^{\dag} =\left(\frac{d}{dr}+W(r)\right)\left(-\frac{d}{dr}+W(r)\right)=-\frac{d^2}{dr^2}+W^2(r)+{dW(r)\over dr}=-\frac{d^2}{dr^2}+U_2(r)
\end{eqnarray}
where the potentials $U_1$ and $U_2$ are known as supersymmetric partner potentials defined by:
\begin{equation}\label{eq17}
U_1(r)=W^{2}(r)-{dW(r)\over dr},\quad \mbox{and}\quad U_2(r)=W^{2}(r)+{dW(r)\over dr}.
\end{equation}
Suppose that $\psi_n^{(1)}(r)$ and $\psi_n^{(2)}(r)$ are the eigenfunctions of $H_1$ and $H_2$, respectively. In the unbroken SUSY case, the ground state is not degenerate with a vanishing energy $E_0=0$ and it is usually expressed in terms of the superpotential $W(r)$ as:
\begin{equation}\label{eq18}
    \mathcal{A} ~\psi^{(1)}_{0}(r)=\left(\frac{d}{dr}+W(r)\right)~\psi^{(1)}_{0}(r)=0\Rightarrow \psi^{(1)}_{0}(r)=C \exp\left(-\int^{r} W(\tau) d\tau\right),
\end{equation}
where $C$ is the normalization constant. The key result is the iso-spectrality between the two Hamiltonians $H_1$ and $H_2$ for all but the ground state ($n = 0$),  which can be shown as follows. Since the energy eigenvalues of $H_1$ and $H_2$ are positive semi-definite $E_n^{1,2}\geq 0$, we have for $H_2\psi_n^{(2)}=\mathcal{A}\mathcal{A}^{\dag}\psi_n^{(2)}=E_n^{(2)}\psi_n^{(2)}$ and by multiplying through by $\mathcal{A}^{\dag}$ we see that
\begin{equation}\label{eq19}
    H_1 (\mathcal{A^{\dag}}\psi^{(2)}_{n})=\mathcal{A}^{\dag}\mathcal{A} (\mathcal{A}^{\dag}\psi^{(2)}_{n})=\mathcal{A}^{\dag}H_{2}\psi^{(2)}_{n}=E^{(2)}_{n}(\mathcal{A^{\dag}}\psi^{(2)}_{n}).
\end{equation}
Thus $A^\dagger\psi_{n}^{(2)}$ is an eigenstate of $H_1$ with same energy eigenvalue $E_n^{(2)}$, and there must be $\psi_n^{(1)}(r)$ such that \begin{equation}\label{eq20}
\psi_n^{(1)}(r)=c_n\mathcal{A^{\dag}}\psi^{(2)}_{n}(r)
\end{equation}
where $c_n$ are constants for $n=1,2,\dots$
Similarly, for $H_1\psi_n^{(1)}=\mathcal{A}^{\dag}\mathcal{A}\psi_n^{(1)}=E_n^{(1)}\psi_n^{(1)}$ and, multiplying through by $\mathcal{A}$, we find
\begin{equation}\label{eq21}
    H_2 (\mathcal{A} \psi^{(1)}_{n})=\mathcal{A} \mathcal{A}^{\dag} (\mathcal{A} \psi^{(1)}_{n})= \mathcal{A} H_1 \psi^{(1)}_{n}=E^{(1)}_{n}(\mathcal{A} \psi^{(1)}_{n}).
\end{equation}
Thus $A\psi_n^{(1)}$ is an eigenstate of $H_2$ with the same eigenvalue $E_n^{(1)}$, and there must be
$\psi_n^{(2)}(r)$ such that
\begin{equation}\label{eq22}
 \psi_n^{(2)}(r)=c_n'\mathcal{A}\psi^{(1)}_{n}(r).
\end{equation}
Furthermore, since
$$1=\langle \psi_n^{(2)}(r)|\psi_n^{(2)}(r)\rangle= |c_n'|^2\langle \mathcal{A}\psi^{(1)}_{n}(r)|\mathcal{A}\psi^{(1)}_{n}(r)\rangle= |c_n'|^2\langle \psi^{(1)}_{n}(r)|\mathcal{A}^\dag\mathcal{A}\psi^{(1)}_{n}(r)\rangle=E_n^1 |c_n'|^2\langle \psi^{(1)}_{n}(r)|\psi^{(1)}_{n}(r)\rangle,$$
we have
\begin{equation}\label{eq23}
 \psi_n^{(2)}(r)=(E_n^{(1)})^{-1/2}\mathcal{A}\psi^{(1)}_{n}(r)=(E_n^{(1)})^{-1/2}\left({d\over dx}+W(r)\right)\psi^{(1)}_{n}(r).
\end{equation}
Similarly, we find
\begin{equation}\label{eq24}
 \psi_n^{(1)}(r)=(E_n^{(2)})^{-1/2}\mathcal{A}^\dag\psi^{(2)}_{n}(r)=(E_n^{(2)})^{-1/2}\left(-{d\over dx}+W(r)\right)\psi^{(2)}_{n}(r).
\end{equation}
Thus, if we knew the eigenvalues and eigenfunctions of either of the two partner potentials, we could immediately
derive the spectrum of the other. However, the above relations only give the
relationship between the eigenvalues and eigenfunctions of the two partner Hamiltonians, but
do not allow us to determine their spectra. In the next, we are guided by the idea of finding pairs of (essentially) isospectral Hamiltonians, one of which has a known soluble Hamiltonian.
\section{Supersymmetric partner of the Isotonic potential}
A key step in this work was the idea to consider a candidate for a superpotential $W(r)$ by means of the following Ansatz:
\begin{equation}\label{eq25}
    W(r)=\frac{k'}{r}+\omega'~ r+\frac{s r}{r^{2}+a^{2}},
\end{equation}
where $k',\omega'$ and $s$ are real parameters to be determine shortly, and $a$ is a fixed potential parameter. The potential $U_1$, then reads
\begin{eqnarray}\label{eq26}
U_1(r)&=&\omega'^2r^2+{k'(k'+1)\over r^2}+{s(-2a^2\omega'+s+2k'+1)r^2-a^2s(2\omega' a^2-2k'+1)\over (r^2+a^2)^2}+(2s+2k'-1)\omega'\nonumber\\
\end{eqnarray}
and the potential $U_2$ is given by
\begin{eqnarray}\label{eq27}
U_2(r)
&=&\omega'^2r^2+{k'(k'-1)\over r^2}+{s(s-1+2k'-2a^2\omega')r^2+a^2s(1+2k'-2a^2\omega')\over (r^2 +a^2)^2}+(2s+2k'+1)\omega'\nonumber\\
\end{eqnarray}
If we now compare the potential $U_1(r)$ with

\begin{equation}\label{eq28}
V_1(r) = \frac{k(k+1)}{r^2} + \omega^2 r^2 + 2g\frac{(r^2-a^2)}{(r^2+a^2)^2}
\end{equation}

\noindent we have for $k'=k$, $\omega'=\omega$, $s(-2a^2\omega'+s+2k+1)=2g$, and $s(2\omega' a^2-2k'+1)=2g$, that
\begin{equation}\label{29}
 \left\{ \begin{array}{l}
 s=4(\omega a^2-k),\\ \\
g=2(wa^2-k)(2wa^2+1-2k).
       \end{array} \right.
\end{equation}
With these values of $s$ and $g$ we may reduce $U_2$ by the assumption that  $s(s-1+2k-2a^2\omega)=0$ and $a^2s(1+2k-2a^2\omega)=0$ to
\begin{equation}\label{eq30}
V_2(r)=\omega^2r^2+{k(k-1)\over r^2}+(8(\omega a^2-k)+2k+1)\omega.
\end{equation}
These assumptions are valid under the following conditions:
$$\omega a^2 =k+{1\over 2},\quad g=2\quad\mbox{and}\quad s=2.$$
In summary, we have two partner potentials
\begin{equation}\label{eq31}
V_{(1)}(r)=\frac{k(k+1)}{r^{2}}+\omega^{2} r^{2}+\frac{4(r^{2}-a^{2})}{(r^{2}+a^{2})^{2}}+\omega(2k+3)
\end{equation}
\begin{equation}\label{eq32}
    V_{2}(r)=\frac{k(k-1)}{r^{2}}+\omega^{2} r^{2}+\omega(2k+5).
\end{equation}
provided the parameters $\omega$ and $a$ satisfy
\begin{equation}\label{eq33}
    \omega a^{2}= k +\frac{1}{2}= \sqrt{B^2 + \left(\ell +\frac{(d-2)}{2}\right)^2}.
\end{equation}
\section{ Gol'dman and Krivchenkov potential in $d$-dimensional}
\noindent Schr\"odinger's equation with the Gol'dman and Krivchenkov potential \cite{hall3} in $d$-dimensions is given by
\begin{equation}\label{eq34}
\left(-{d^2\over dr^2}+{\Lambda(\Lambda+1)\over r^2} +\beta r^2+{\alpha\over r^2}\right)\psi_{n\gamma_d}(r)=E_{n\gamma_d}\psi_{n\gamma_d}(r),\quad \Lambda=l+{1\over 2}(d-3),~~ d\geq 2,
\end{equation}
and has exact eigenvalues given by \cite{hall3}
\begin{equation}\label{eq35}
E_{n\gamma_d}=2\sqrt{\beta}\left(2n+\gamma_d\right),\quad n,l=0,1,2,\dots,
\end{equation}
where
\[
 \gamma_d=1+\sqrt{\alpha+(\Lambda+{1\over 2})^2}.
\]
Meanwhile, the exact eigenfunctions are given by \cite{hall3}
\begin{equation}\label{eq36}
\psi_{n\gamma_d}(r)=(-1)^n\sqrt{2\beta^{\gamma_d/2}(\gamma_d)_n\over n!\Gamma(\gamma_d)}r^{\gamma_d-1/2}e^{-{1\over 2}\sqrt{\beta}r^2}{}_1F_1(-n;\gamma_d;\sqrt{\beta}r^2).
\end{equation}
Here, we use the Pochhammer symbol $(\gamma)_n$,  where
$$(\gamma)_n=\gamma(\gamma+1)(\gamma+2)\dots(\gamma+n-1)={\Gamma(\gamma+n)\over \Gamma(\gamma)},$$
and ${}_1F_1$ is the confluent hypergeometric function defined by (\cite{rainville}, chapter 7)
\begin{equation}\label{eq37}
{}_1 F_1(-n;b;x)=\sum_{k=0}^n {(-n)_k\over (b)_k~k!} x^k.
\end{equation}
Using these exact solutions for the Gol'dman and Krivchenkov potential, we can show that the Schr\"odinger equation
\begin{equation}\label{eq38}
\left(-{d^2\over dr^2}+{k(k-1)\over r^2} +w^2 r^2+\omega(2k+5)\right)\psi_{nk}(r)=\epsilon_{nk}\psi_{nk}(r)
\end{equation}
has exact solutions given by
\begin{equation}\label{eq39}
\epsilon_{nk}=2\omega (2n+2 k+3),
\end{equation}
and
\begin{equation}\label{eq40}
\psi_{nk}(r)=(-1)^n\sqrt{2\omega^{k+1/2}(k+1/2)_n\over n!\Gamma(k+1/2)}r^{k}e^{-{1\over 2}\omega r^2}{}_1F_1(-n;k+{1\over 2};\omega r^2),
\end{equation}
where
\begin{equation}\label{eq41}
k +\half = \gamma_d = \omega a^2 = \left[B^2 +\left(\ell + \frac{d-2}{2}\right)^2\right]^{\half}.
\end{equation}
It is clear from these relations that states having the same value for the combination $2\ell +d$ are degenerate.
\section{Exact solutions for isotonic nonlinear-oscillator potentials}
The supersymmetric approach, briefly discussed in section~3, along with exact solution of the Gol'dman and Krivchenkov potential allow us to obtain the exact solutions of Schr\"odinger's equation
\begin{equation}\label{eq42}
\left(-{d^2\over dr^2}+\frac{k(k+1)}{r^{2}}+\omega^{2} r^{2}+\frac{4(r^{2}-a^{2})}{(r^{2}+a^{2})^{2}}+\omega(2k+3)\right)\phi_{nk}(r)=\epsilon_{nk}\phi_{k}(r),
\end{equation}
namely,
\begin{equation}\label{eq43}
\epsilon_{nk}=2\omega (2n+2 k+3),
\end{equation}
and the corresponding wavefunctions are given by
\begin{eqnarray}\label{eq44}
\phi_{nk}(r)&=&C_n'\left(-{d\over dr}+{k\over r}+\omega r+{2r\over r^2+a^2}\right)r^{k}e^{-{1\over 2}\omega r^2}L_{n}^{k-{1\over 2}}(\omega r^2)
\end{eqnarray}
where
\begin{eqnarray}\label{eq45}
C'=(-1)^n\sqrt{2\omega^{k+1/2}(k+1/2)_n\over 2\omega (2n+2 k+3)~n!\Gamma(k+1/2)}.
\end{eqnarray}
Here we have used the well-known relation between the confluent hypergeometric function (\ref{eq37}) and Laguerre polynomials (\cite{rainville}, page 203)
\begin{equation*}
{}_1F_1(-n;\alpha+1;z)={n!\over (\alpha+1)_n}L_n^{\alpha}(z).
\end{equation*}
A straightforward computation, aided by the differential identity of Laguerre polynomials (\cite{rainville}, page 203, formula (11)), namely
\begin{eqnarray}\label{eq46}
{d\over dz}L_n^{\alpha}(z)=-L_{n-1}^{\alpha+1}(z),
\end{eqnarray}
yields
\begin{eqnarray}\label{eq47}
\phi_{nk}(r)&=&{C_n' r^{k+1}e^{-\omega r^2/2}\over r^2+a^2}\bigg[(2k+2n+3)L_n^{k-1/2}(\omega r^2)-2(n+1)L_{n+1}^{k-1/2}(\omega r^2)+2\omega a^2L_n^{k+1/2}(\omega r^2)\bigg],\nonumber\\
\end{eqnarray}
where we have used the identity (\cite{rainville}, page 203, formula (8))
\begin{eqnarray}\label{eq48}
L_n^{\alpha}(z)=L_{n-1}^\alpha (z)+L_n^{\alpha-1}(z).
\end{eqnarray}
Since $wa^2=k+1/2$,  we may write Eq.(\ref{eq47}) as
\begin{eqnarray}\label{eq49}
\phi_{nk}(r)&=&{C_n' r^{k+1}e^{-\omega r^2/2}\over r^2+a^2}\bigg[(2k+2n+3)L_n^{k-1/2}(\omega r^2)-2(n+1)L_{n+1}^{k-1/2}(\omega r^2)+(2k+1)L_n^{k+1/2}(\omega r^2)\bigg].\nonumber\\
\end{eqnarray}
This may be compared with the results of Fellow and Smith \cite{fellows} by use of the well-known relations between Hermite and Laguerre polynomials (\cite{rainville}, page 216, problem (1))
\begin{equation}\label{eq50}
    L^{-1/2}_{n}(\omega r^{2})=\frac{(-1)^{n}}{2^{2n} n!} H_{2n}(\sqrt{\omega} r)
\end{equation}
and
\begin{equation}\label{eq51}
    L^{1/2}_{n}(\omega r^{2})=\frac{(-1)^{n}}{2^{2n+1} n!~\sqrt{\omega}~ r} H_{2n+1}(\sqrt{\omega} r).
\end{equation}
For $k=0$, $\omega=1$ using the identity (\cite{rainville}, page 188, formulas (4) and (6))
\begin{eqnarray}\label{eq52}
2zH_k(z)=H_{k+1}(z)+2kH_{k-1}(z).
\end{eqnarray}
we obtain
\begin{eqnarray}\label{eq53}
\phi_{nk}(r)&=&{(-1)^n C_n' e^{-r^2/2}\over 2^{2n+2}~n!~(r^2+a^2)}\bigg[4(2n+3)H_{2n+1}(r)+8n(2n+3)H_{2n-1}(r)+H_{2n+3}( r)\bigg],
\end{eqnarray}
where
\begin{eqnarray}\label{eq54}
C'=(-1)^n\sqrt{\Gamma(n+{1\over 2})\over (2n+3)~n!~\pi}.
\end{eqnarray}
These general results agree with the results discussed by  Cari\~{n}ena et al \cite{perelomov}, and Fellows et al. \cite{fellows} for the comparable odd solutions $n=3,5,7,\dots$ in Eq.(\ref{eq2}).
\section{Conclusion}
In this paper, we have studied a family of generalized quantum nonlinear oscillators in $d$-dimensions. These isotonic oscillator problems have potentials $V(r)$ of the form
\begin{equation}\label{eq55}
V(r) = \frac{B^2}{r^2} + \omega^2 r^2 + 2g\frac{(r^2-a^2)}{(r^2+a^2)^2}.
\end{equation}
We show that if $g=2$ and $(\omega a^2)^2 = B^2 +(\ell +(d-2)/2)^2,$ then this potential can be regarded as a supersymmetric partner  of the Gol'dman and Krivchenkov potential, for which exact solutions can be constructed.  Thus, under these conditions, we are able to solve the eigenproblem $H\psi = (-\Delta + V)\psi = E\psi$ exactly in $d$ dimensions, and provide formulae for all the discrete eigenvalues and corresponding normalized wave functions. 
 We have shown that our solutions to the general problem agree with the results reported earlier in Refs. \cite{perelomov} and \cite{fellows} for the comparable odd states $n=3,5,7,\dots$ in $d=1$ dimension with $B = 0.$ 
\medskip
\section*{Acknowledgements}
Partial financial support of this work under Grant Nos. GP3438 and GP249507 from the
Natural Sciences and Engineering Research Council of Canada
 is gratefully acknowledged by two of us (RLH and NS).
Two of us (\"OY and NS) would also like
to thank the Department of Mathematics and Statistics of Concordia University
 for its warm hospitality.

\medskip




\begin{thebibliography}{99}
\bibitem{perelomov} J. F. Cari\~{n}ena, A. M. Perelomov, M. F. Ra\~{n}ada and M. Santander, J. Phys. A: Math. Theor. \textbf{41} 085301 (2008).
\bibitem{fellows} J. M. Fellows and R. A. Smith, J. Phys. A: Math. Theor. \textbf{42} 335303 (2009).
\bibitem{sesma} J. Sesma, J. Phys. A: Math. Theor. \textbf{43}  185303 (2010).
\bibitem{kra} R. A. Kraenkel and M. Senthilvelan, J. Phys. A: Math. Theor. \textbf{42} 415303 (2009).
\bibitem{cooper} F. Cooper, A. Khare, U. Sukhatme, Phys. Rep. \textbf{251} 268 (1995).
\bibitem{hall2} R. L. Hall and \"{O}. Ye\c{s}ilta\c{s}, Int. J. Mod. Phys. E (In Press);  {\tt arXiv:math-ph/1006.4628}.
\bibitem{sommerfeld} A. Sommerfeld,{\it Partial Differential Equations} (New York, Academic Press, 1949).
\bibitem{hall1} R. L. Hall, Q. D. Katatbeh, N. Saad, J. Phys. A: Math. Gen. \textbf{37} 11629 (2004).
\bibitem{hall3} R. L. Hall, N. Saad, and A. B. von Keviczky,
J. Math. Phys. 43, 94-112 (2002).
\bibitem{rainville} E. D. Rainville, \emph{Special functions}, Chelsea Publishing Company, New York, 1960.

\end{thebibliography}
\end{document}